# Structural and optical characterization of NiO polycrystalline thin films fabricated by spray-pyrolysis

**Lakshmi Das, Esdras J. Canto-Aguilar, Tlek Tapani, Haifeng Lin, Hinduja Bhuvanendran, Nicolas Boulanger, Roushdey Salh, Eduardo Gracia-Espino, Nicolò Maccaferri**

[1] Department of Physics, Umeå University, Umeå, Sweden

E-mail: nicolo.maccaferri@umu.se; lakshmi.thazhe@umu.se

**Abstract**
Nickel (II) oxide, NiO, a wide band gap Mott insulator characterized by strong Coulomb repulsion between d-electrons and displaying antiferromagnetic order at room temperature, has gained attention in recent years as a very promising candidate for applications in a broad set of areas, including chemistry and metallurgy to spintronics and energy harvesting. Here, we report on the synthesis of polycrystalline NiO fabricated using spray-pyrolysis technique, which is a deposition technique able to produce quite uniform films of pure and crystalline materials without the need of high vacuum or inert atmospheres. We then characterized the composition and structure of our NiO thin films using X-ray diffraction, and atomic force and scanning electron microscopies, respectively. We completed our study by looking at the phononic and magnonic properties of our NiO thin films via Raman spectroscopy, and at the ultrafast electron dynamics by using optical pump probe spectroscopy. We found that our NiO samples display the same phononic and magnonic dispersion expected for single crystal NiO at room temperature, and that electron dynamics in our system is similar to those of previously reported NiO mono- and poli-crystalline systems synthesized with different techniques. These results prove that spray-pyrolysis can be used as affordable and large-scale fabrication technique to synthetize strongly correlated materials for a large set of applications.

## 1. Introduction

The burgeoning interest in miniaturisation of technological devices and reducing energy dissipation has led to a pursuit of exploring materials such as metal oxides for a broad range of applications spanning from energy and information storage to metallurgy and advanced manufacturing. Within the family of metal oxides, strongly correlated materials such as Mott insulators, in particular those displaying antiferromagnetic order, have received particular attention. Several features of antiferromagnetic materials like net zero magnetization, absence of stray fields, robustness against external magnetic perturbations and ultrafast magnetization dynamics in the THz range make them promising systems in several applications, for instance spintronics [1], [2], [3], [4]. In this context, one very promising material is NiO, which is a prototypical wide band gap (4.3 eV) [5] antiferromagnet (AFM) with a Néel transition temperature at 523K [6]. It has a cubic rock salt structure with $Ni^{2+}$ magnetic moments aligned ferromagnetically within {111} planes and exhibits AFM coupling between the {111} planes [7]. It belongs to the family of strongly correlated electron systems and is classified as an intermediate charge transfer insulator as per the Zaanen - Sawatsky - Allen (ZSA) classification scheme [7]. Several works have reported using NiO as a potential candidate material in the field of antiferromagnetic spintronics where the moments are manipulated electrically or optically to produce fast response of the spin system to the applied external stimuli [8], [9], [10], [11]. However, such works require precise and controlled synthesis of well-ordered single crystalline AFM materials to study the transport properties and the magnetization dynamics.

The synthesis of NiO particles and films has been addressed with different physical and chemical techniques like laser ablation, chemical vapor deposition, sputtering, sol-gel, microwave-assisted reduction, co-precipitation and spray-pyrolysis, among others [12]. For practical applications, deposition techniques able to produce uniform films of pure and crystalline materials without the need of high vacuum or inert atmospheres, are important for large-scale fabrication of devices. Spray-pyrolysis (SP) has already met satisfactorily these requirements [13]. For over four decades, SP has been considered the choice of excellence for the synthesis of different metal oxides widely used in photovoltaics [14], [15], [16] batteries and supercapacitors [17], [18], catalyst [19], [20], etc.; due to the possibility to easily synthetize these materials as powders or films and with different morphologies by varying deposition parameters or the precursors solution composition, depending on the needs [21]. The deposition of NiO thin films by SP can be done through different Ni-based precursors (as acetates, nitrates, sulfides, hydroxides), where chlorides show a direct decomposition mechanism which usually leads to crystalline and uniform films [22], [23].

In this work we report the synthesis and characterization of polycrystalline NiO films prepared via SP (NiO polycrystal) of ethanol-based $NiCl_2$ solutions, as a cost-effective and large-scale deposition technique of high-quality thin films. We characterize the topography of the samples using Atomic Force microscopy (AFM) and Scanning Electron Microscopy (SEM). We also make use of optical spectroscopy techniques such as X-ray Diffraction (XRD) as well as Raman and pump-probe spectroscopy to characterize the structural, electronic and crystal properties of our samples. We found that such properties are similar to those of previously reported NiO mono- and poli-crystalline systems synthesized with other techniques. Thus, this proves that SP can be used as affordable and large-scale fabrication technique to synthetize strongly correlated materials for a large set of applications.

## 2. Synthesis of NiO thin films

The deposition of NiO thin films by SP was carried with a solution 0.1M $NiCl_2 \bullet 6H_2O$ (Sigma-Aldrich, ≥ 95 %) in anhydrous ethanol (VWR, 100 %), a growth temperature of 500 °C. Compressed air was used as carrier gas at an outlet pressure of 2 bar, as shown in Figure 1.

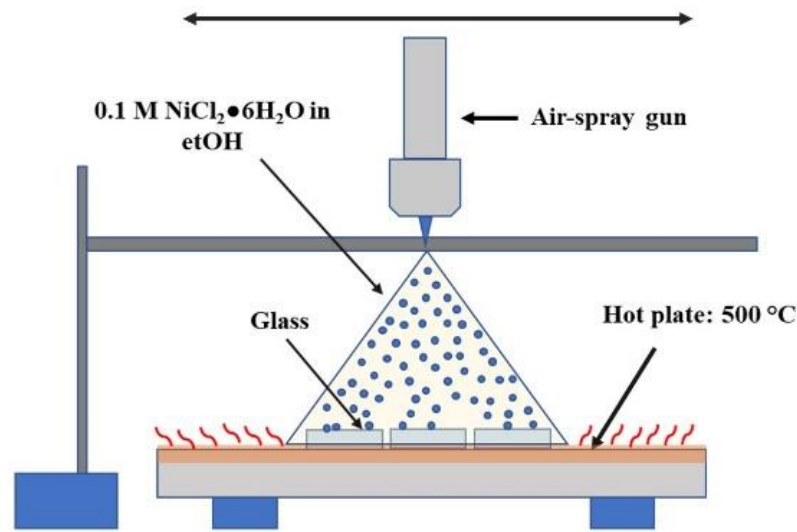

**Figure 1**: **Schematic of the film growth technique;** Schematic illustration of the spray-pyrolysis setup used for the synthesis of NiO polycrystal.

The desired amount of the precursors solution (100 or 200 ml, depending on the sample) was sprayed onto the hot substrates (glass and Fluorinated Tin oxide (FTO) ) by sweeping the air-spray gun from side to side of the hotplate (19.5 cm of length) at a speed of ≈ 4.9 cm s$^{-1}$ and a feeding flow rate of ≈0.068 mL s$^{-1}$. After that, the sample was kept for 1 h more on the hotplate at 500 °C and cooled down at room temperature. Finally, a thermal treatment at 500 °C for 2 h in a furnace was applied to ensure a complete conversion of remnant reactants into pure and crystalline NiO.

The mechanism of formation of crystalline NiO through the pyrolysis of NiCl$_2$, according to the literature [22], [23], is:

$$NiCl_2 \bullet 6H_2O \rightarrow NiO + 2HCl\uparrow + 5H_2O\uparrow \quad (1)$$

## 3. Characterization

**3.1 Atomic Force Microscopy and Scanning Electron Microscopy -** AFM measurements were carried out in non-contact mode to characterize the morphology of the deposited film on glass substrate. Figure 2 shows the AFM and profilometer measurements on the sample. The AFM image shows a rough and porous sample surface with average grain size of ~1.2 μm estimated from a 10×10 μm scan area. Average roughness along an arbitrary horizontal line drawn across the AFM scan is estimated to be ~ 60 nm using Gwyddion Software. The thickness of the sample has been estimated using a Bruker Dektak XT profilometer to be 2.5 ± 0.02 μm.

SEM at magnifications 1KX, 25KX, 50KX, 100KX were also used to analyse the morphology of the sample grown on FTO (Fluorine Tin oxide coated glass substrate) which is a transparent conductive oxide coated glass as SEM imaging requires samples on conductive surfaces, as shown in Figure 3. The films displayed an almost flat and slightly rough surface due to the compact arrangement of small NiO particles and the presence of agglomerates of sharp-larger particles, thus complementing the AFM results, which could not resolve the real dimensions of the elements within the agglomerates with size of ~1.2 μm.

**3.2 X-ray Diffraction (XRD) -** The thin film sample of NiO deposited on glass substrate was characterized using X-ray diffraction using a PANalytical Xpert3 Powder XRD. X-ray diffraction pattern obtained for NiO thin films of thickness ~ 3 μm is shown in Figure 4. The presence of several peaks in the sample across a wide range of 2θ indicates the polycrystalline nature of the sample is cubic phase with observed XRD peaks corresponding to the planes (111) and (002), respectively, and with a preferred orientation in the (111) direction. The presence of secondary peaks corresponding to impurities (NaCl) indicate a side reaction between Na$_2$O present in glass and the precursor solution. For comparison, XRD pattern for a single crystalline NiO in (111) direction is also shown in Fig 4.

**3.3 Raman spectroscopy -** Figure 5 displays the Raman spectra obtained for NiO polycrystalline samples of thickness ~ 3 μm at room temperature in the wavenumber range 400-2000 cm$^{-1}$. Raman spectroscopy measurements were carried out in a Renishaw Qontor Raman spectrometer with laser excitation wavelength of 405 nm. The Raman spectra shows several features. The first three peaks are vibrational, and the peak at 560 cm$^{-1}$ is a 1st order phonon mode, and the peaks at 780cm$^{-1}$ and 1100 cm$^{-1}$ are assigned to 2nd order phonon modes. Due to its high $T_N$, NiO is AFM at roo temperature, and the small peak around 1600 cm$^{-1}$ can be attributed to the two magnon 2M mode [24], [25], [26].

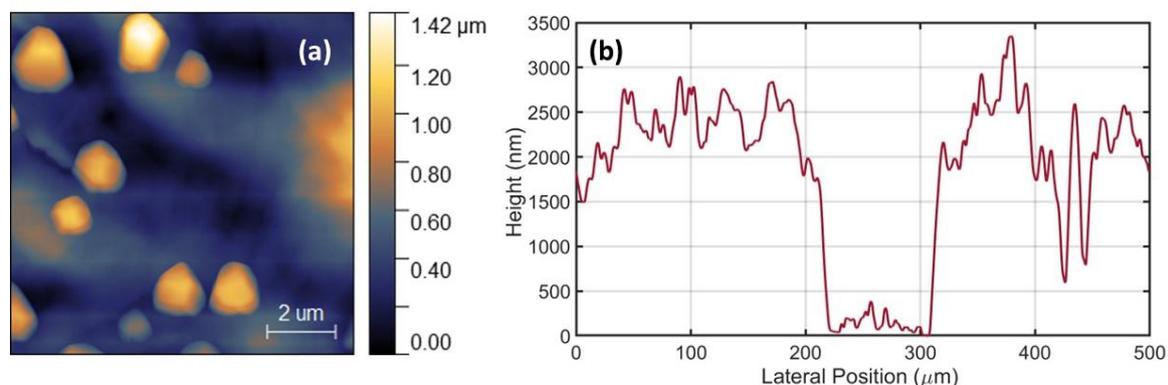

**Figure 2: AFM and profilometer measurements;** (a) AFM scan of the polycrystal NiO on glass substrate. (b) The film thickness is estimated to be around 2.5 μm from profilometer measurements.

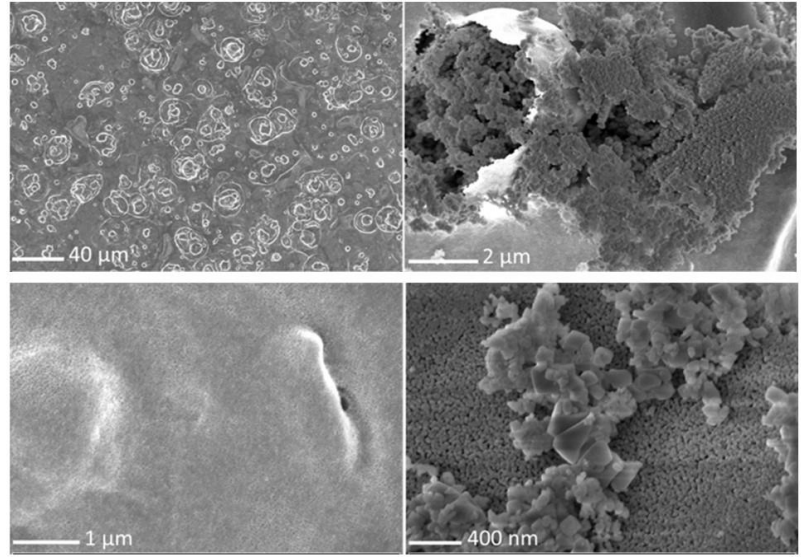

**Figure 3: SEM**; SEM micrographs (at different magnifications) of polycrystalline NiO thin films deposited on FTO substrates.

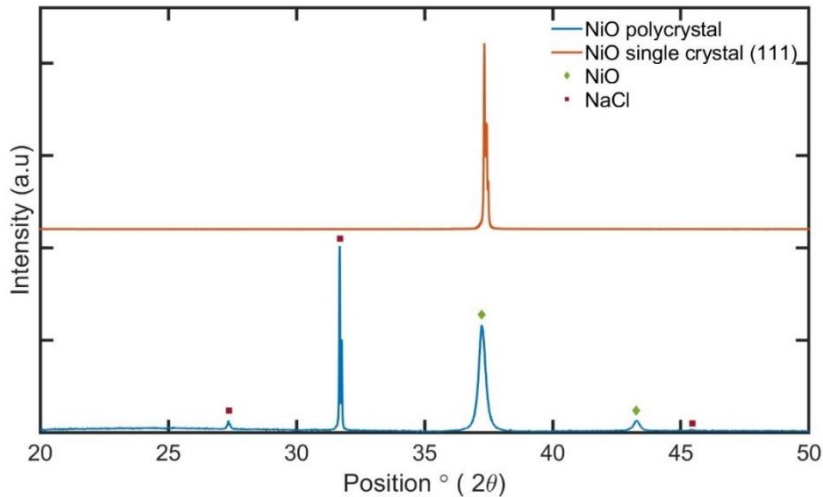

**Figure 4: XRD patterns**; XRD patterns obtained for NiO polycrystalline thins film and a preferred (111) oriented single crystal (blue spectrum). The XRD for single crystal (orange spectrum) is shifted arbitrarily along vertical y- axis for clarity.

### 3.4 Ultrafast pump probe measurements

Pump-probe optical spectroscopy measurements using sub-15 fs light pulses were carried out to study the carrier dynamics in the NiO polycrystalline thin film samples. For our measurements we have employed a Yb:KGW amplifier laser (operational wavelength of 1030nm, 50 KHz repetition rate and average output power of 20 W) pumping two individual optical parametric amplifiers (OPAs). The outputs generated by the OPAs served as the pump and probe pulses in the measurements. These pulses have center wavelengths of 850nm (1.5 eV) and 620nm (2 eV), respectively, as depicted in Figure 6 (a). To resolve the dynamics of the system in the time domain, time delay between the pump and probe was controlled with sub-fs precision by a PI L-511 High-Precision Linear Stage placed along the pump beam path. In our experiments, we have measured the transient reflectivity, that is the pump-induced reflectivity changes defined as

$$\frac{\Delta R}{R} = \frac{R_t - R_0}{R(0)} \qquad (2)$$

where $R_t$ and $R_0$ denote the reflectivity of the system after excitation and at a certain time delay between the pump and the probe pulses, and at ground state (pump pulse is not exciting the system), respectively.

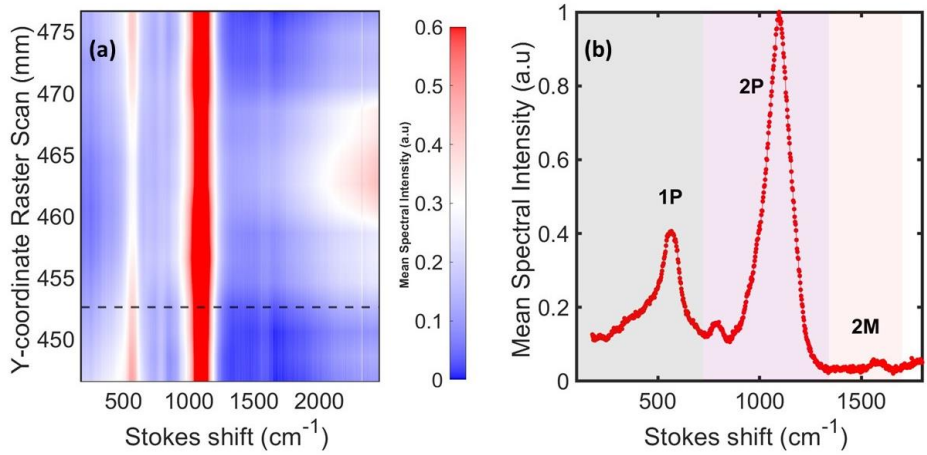

**Figure 5: Raman spectra of the NiO polycrystal at RT**; (a) 2D colour plot of the Raman spectra (raster scan over the sample area) with mean spectral intensity across a scan on the sample. The horizontal black dashed line indicates an arbitrary line scan along a specific spatial direction on the sample. (b) Raman spectra for the line scan at the position indicated by the black dashed line in (a). The peaks at 500-700 cm$^{-1}$ and the range 700-1200 cm$^{-1}$ indicate the first order and the second order phonon modes, respectively, while the peak around 1600 cm$^{-1}$ represents the 2-magnon mode.

Figure 6(a) shows the spectra of the pump and probe pulses, employed in our measurements, that perturbs and measure the optical response of the system respectively. Figure 6(b) and (c) shows the transient reflectivity measurements of the system under investigation, a NiO policrystalline thin film of thickness ~ 3 μm, pumped and probed by pulses centered at 850 nm and 620nm respectively. Figure 6(b) shows a 2D color map of the transient reflectivity of the sample as a function of wavelength (y-axis) and time delay between the pump and the probe pulses (Δt, x-axis). The spectra obtained from the colorplot at a wavelength 600 nm (indicated by a black dotted line) is shown in Fig 6 (c). The transient reflectivity traces exhibit negative differential reflectivity and shows an initial drop followed by a relaxation process where the system recovers its equilibrium state. At negative time delays (probe arriving before the pump pulse), no changes in reflectivity can be identified as shown in Figure 6 (c). The time t=0 indicates when the pump and the probe arrive at the system simultaneously. Here, the spectra show a sharp drop due to decrease in reflectivity. NiO is an intermediate charge transfer insulator with oxygen *p*-band located near the Lower Hubbard Band (LHB) energy range [7]. Optical excitation in charge transfer insulators can lead to modification of the Mott Hubbard band gap where the band gap between upper and lower Hubbard bands are narrowed [27]. Thus the negative signal ΔR/R after non resonant excitation below band gap could arise from photo-induced band gap renormalization in NiO [28]. After the initial exponential drop, the system relaxes back to its equilibrium state at longer time delays due to energy redistribution arising from electron-electron and electron-phonon scattering. The reflectivity dynamics can be phenomenologically modelled by the exponential function:

$$M(t) = A_1 exp\frac{-(t-t_0)}{\tau_{fast}} + C \qquad (3)$$

Figure 6(c) shows the exponential fit on the transient reflectivity spectrum and the relevant time scale $\tau_{fast}$ characterizing the fast relaxation dynamics has been estimated to be around 104±7 fs, which is comparable to previous results obtained in similar NiO nanoclusters [29]. Figure 7 (a), (b), (c) and (d) displays the measured transient reflectivity changes and the corresponding fits as a function of varying pump fluences. The fluence was estimated from gaussian fitting the beam profile, which is reasonable if considering that we are using few cycle light pulses. Figure 7(e) shows the fluence dependence of the absolute value of the maximum rise amplitude. The peak amplitude shows a monotonous increase with increase in fluence indicating that more carriers are excited when pump fluence is increased. The fast relaxation time constant $\tau_{fast}$ was analysed as a function of the pump fluence (see Figure 7(f)). The calculated fast time constant remained around ~ 127 fs (average) for the applied laser fluences. Such short relaxation times has been attributed in previous studies to fast direct recombination of photoexcited electrons and holes due to reduced screening of Coulomb interaction [30].

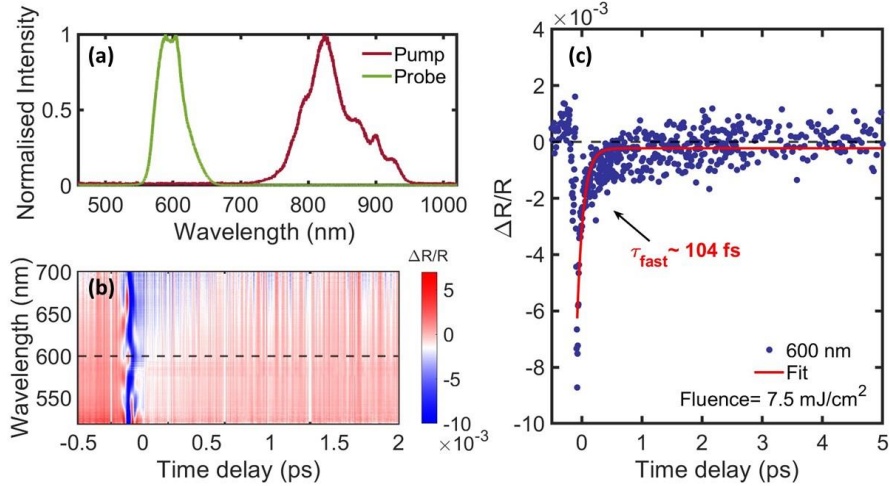

**Figure 6: Pump-probe measurements to study charge dynamics;** (a) Normalized spectra of the pump (fluence 7.5 mJ/cm$^2$) and probe pulses used in the measurements. (b) 2D color map of the reflectivity changes as a function of wavelength and time delay. The black dashed line indicates a representative wavelength where the signal is analyzed. (c) Time dependence of transient reflectivity for the NiO polycrystalline thin film of thickness ~ 3 μm at 600 nm for an applied laser fluence of 7.5 mJ/cm$^2$. The figure shows the data and the fit used (from equation (3)) to estimate the time relaxation constant.

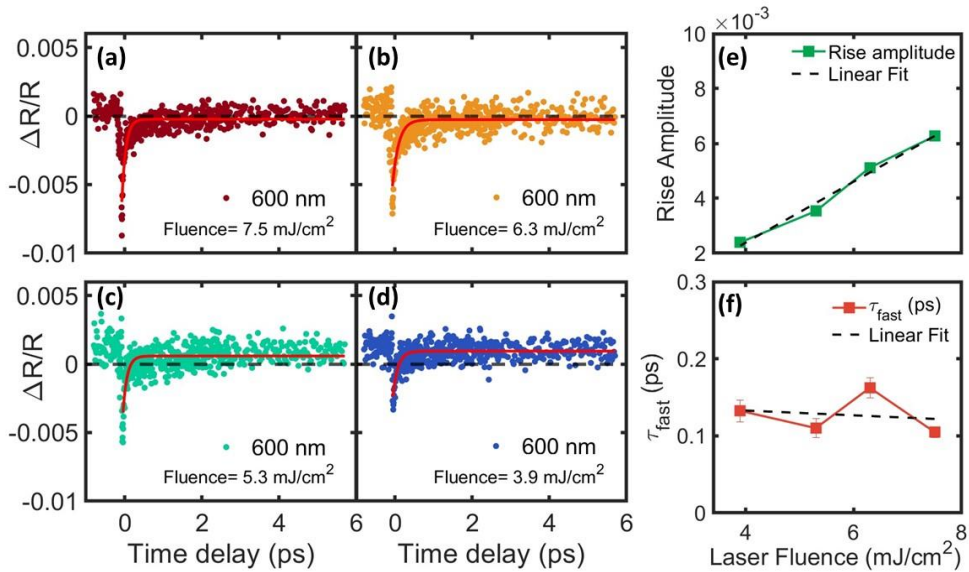

**Figure 7: Fluence dependence of the transient reflectivity.** (a), (b), (c) and (d) shows the transient reflectivity traces with applied laser fluences of 7.5, 6.3, 5.3 and 3.9 mJ/cm$^2$, respectively. The red lines are exponential fits calculated using Eq. (3). (e) and (f) shows the fluence dependence of the absolute value of maximum signal amplitude and the time constants $\tau_{fast}$ obtained from the fit, respectively.

In conclusion, we showed that NiO policrystalline thin films can be synthetize using an affordable and easy technique such as spray-pyrolysis, which is a able to produce quite uniform films of pure and crystalline materials without the need of high vacuum or inert atmospheres. We then characterized the composition and structure of the NiO thin films using X-ray diffraction, and atomic force and scanning electron microscopies, respectively. Noticeably, XRD characterization revealed the presence of polycrystalline NiO with strong (111) orientation, and room temperatyre Raman measurements indicated the presence of both phononic and antiferromagnetic magnon peaks. Transient reflectivity measurements based on non-degenerate pump-probe spectroscopy technique were also conducted to study the charge dynamics of NiO polycrystalline thin films. The fast electronic relaxation time was shown to be around 127 fs. Such short relaxation times has been attributed in previous studies to time constant for photoexcited electron hole direct recombination [30]. Overall, electron dynamics in our system is similar to those of previously reported NiO mono- and poli-crystalline systems synthesized with different techniques, thus proving that spray-pyrolysis can be used as affordable and large-scale fabrication technique to synthetize strongly correlated materials for a large set of applications. To gain further understanding about other properties of the system, such as the possible manipulation of the AFM order, the next step is to measure its magneto-optical response under perturbation of fs light pulses, that could shed information about the spin dynamics of this policrystalline strongly correlated system, and compare this to the ultrafast spin dynamics properties of a monocrystalline NiO thin film.

**Acknowledgements**
LD and NM acknowledge support from the Wenner-Gren Foundation (grant nr: UPD2022-0074). NM, TP, HL, HB acknowledge support from the Swedish Research Council (grant no. 2021-05784), Kempestiftelserna (grant no. JCK-3122) and the European Innovation Council (grant n. 101046920). EJC acknowledges support from the Carl Tryggers Foundation (CTS 21-1581). EGE acknowledges support from Vetenskapsrådet (2018-03937), SSF-Agenda 2030 – PUSH, the Kempe Foundation (JCK-2132), and the Carl Tryggers Foundation (CTS 21-1581). We also thank the Umeå Core Facility for Electron Microscopy (UCEM), the Vibrational Spectroscopy Core Facility (ViSp), and the Nanolab platform at Umeå University.